\documentclass{PoS}
\usepackage{epsfig}
\usepackage{amssymb, amsmath, mathbbol}

\title{Synergy and complementarity between neutrino physics and 
low-energy intensity frontiers}

\ShortTitle{Neutrino physics and the low-energy intensity frontiers}

\author{\speaker{Ana M. Teixeira}\\
        Laboratoire de Physique de Clermont, CNRS/IN2P3 -- UMR 6533,\\ 
        Campus des C\'ezeaux, 4 Av. Blaise Pascal, 
        F-63178 Aubi\`ere Cedex, France\\
        E-mail: \email{ana.teixeira@clermont.in2p3.fr}}

\abstract{Massive neutrinos and leptonic mixings have provided
  the first evidence of flavour violation in the lepton sector,
  opening a unique gateway to many new phenomena. Among the latter,
  one finds processes violating lepton number, charged lepton flavours,
  or even the universality of lepton flavours. These very rare
  transitions can be studied in high-intensity facilities, and if
  observed, constitute a clear sign of New Physics.  
  After a brief review of the experimental status of dedicated
  searches, we comment on the prospects of several well-motivated
  models of neutrino mass generation to several of the above mentioned
  observables, also discussing how the interplay of high-intensity
  observables and neutrino data can shed light on the underlying New
  Physics model.} 

\FullConference{The 19th International Workshop on Neutrinos from
  Accelerators-NUFACT2017\\ 
		25-30 September, 2017\\
		Uppsala University, Uppsala, Sweden}

\begin{document}

\section{Introduction}

Neutrino oscillations provided the first confirmation that the
Standard Model (SM) should be extended; eversince, 
the quest for the new physics (NP)
model accounting for neutrino masses and mixings has become one of the
most active quests in particle physics. There is a vast array
of well-motivated SM extensions, relying on additional fields,
extended gauge groups, or even complete NP frameworks - all capable of
successfully accommodating neutrino oscillation data. 
Fortunately, massive neutrinos and leptonic mixings 
offer a true gateway to many
experimental signals that are either forbidden or extremely suppressed
in the SM; these include charged lepton flavour violation (cLFV),
lepton number violation (LNV), contributions to lepton dipole moments,
among many others. The interplay of oscillation data with the results
of the searches for these rare processes, which are being actively
looked for at numerous high-intensity facilities, may then allow to shed
light on, and hopefully identify, 
the underlying model of neutrino masses and mixings. 

In the original formulation of the SM, neutrinos are strictly massless
and lepton number(s) are conserved; leptonic electric dipole moments
(EDMs) are generated at 4-loop level, and their value is tiny
($d_e^\text{CKM} \leq 10^{-38}e$cm). Minimal SM extensions in which
Dirac $\nu$ masses are put by hand, and the leptonic mixing matrix
$U_\text{PMNS}$ accounts for $\nu$ oscillation data, still preserve
total lepton number; cLFV transitions can theoretically occur, but the
smallness of neutrino masses strongly suppresses the branching
fractions, rendering them unobservable. Likewise, and despite being
generated at the 2-loop level, EDMs still remain beyond experimental
sensitivity.    

Reviews of the experimental status of numerous (rare) leptonic
processes were conducted in the dedicated NUFACT2017 
sessions, as
well as in plenary presentations~\cite{NUFACT2017}; a full summary of current
bounds, including electric and magnetic (anomalous) lepton moments can
be found in~\cite{Patrignani:2016xqp}. 

A detailed discussion of the model-independent approach to constrain
NP models based on the searches for the above mentioned
observables has been done in~\cite{EFT:nufact17}; in what follows,
we focus the discussion on specific NP realisations - from simple,
minimal SM extensions, to complete frameworks.

\section{SM extensions via sterile neutrinos}  
Sterile fermions are an integral part of several well-motivated
mechanisms of neutrino mass generation; before addressing the
contributions of the latter constructions to different high-intensity
observables (and discuss how the interplay of distinct signals can
favour or exclude them), a first phenomenological - and convenient -
approach consists in considering simple ``toy models'', in which the
SM is extended by a single massive sterile state (possibly encoding the
effects of a larger number of sterile states).

\subsection{Minimal ``3+1 toy model''}
Without any assumption on the mechanism of neutrino mass generation,
this simple construction relies in extending the SM content via one
massive heavy sterile state; the interaction and physical basis are
related via a $4\times4$ unitary matrix $U$ (whose upper $3\times
3$ block encodes left-handed leptonic mixings). The non-negligible
active-sterile mixings are at the source of modified charged and
neutral lepton currents, and hence of new contributions to many
observables\footnote{In many of the subsequent numerical results, 
the additional physical parameters of the model were randomly sampled
from the following intervals: $m_4 \in [0.1 - 10^5]$~GeV,
$0\lesssim \theta_{\alpha 4}\lesssim 2 \pi$ (and likewise for the 
CP violating phases).}. 

\begin{figure}[h!]
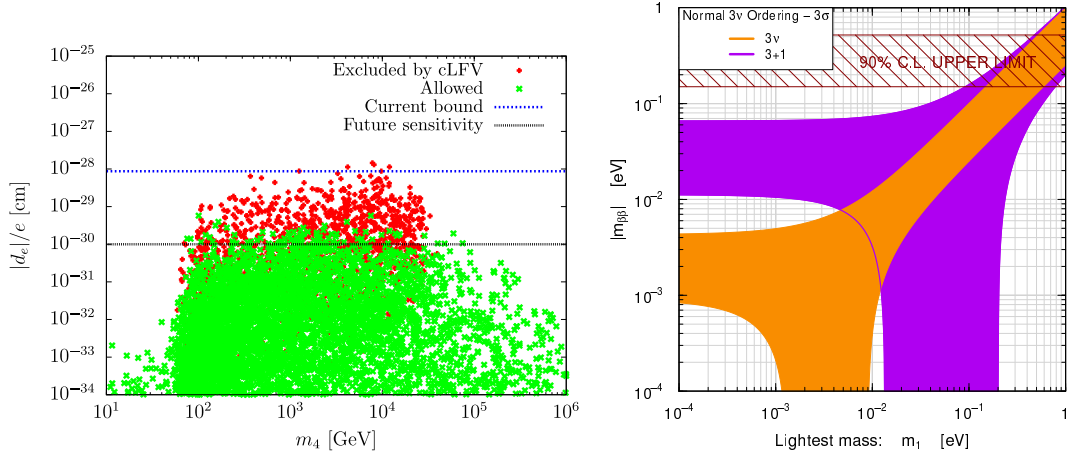

\begin{tabular}{cc}
\epsfig{file=edm_e_m4.epsi, width=75mm} 
&
\epsfig{file=fig-05a.epsi, width=60mm}
\end{tabular}
\caption{On the left, contributions to $|d_e|/e$ as a function of the sterile
  neutrino masses; from~\cite{Abada:2015trh}. 
On the right, effective Majorana mass $m_{\beta \beta}$ as a function of 
the lightest (active) neutrino mass (normal ordering of the light
$\nu$ spectrum); from~\cite{Giunti:2015wnd}. 
}\label{fig:EDM.0nu2beta}
\end{figure}

For example, 
concerning EDMs, the new Majorana and Dirac phases induce non-vanishing
contributions; in the presence of two non-degenerate sterile states
(with masses between 100~GeV and 100~TeV), one can have $|d_e|/e \geq
10^{-30}$~cm, within future ACME sensitivity~\cite{Abada:2015trh}, as
displayed on the left panel of Fig.~\ref{fig:EDM.0nu2beta}. 
Since the Majorana contribution is dominant, the interpretation of an
EDM observation in such a minimal framework would suggest CP
violating Majorana neutrinos, with potential implications for
leptogenesis.   

\begin{figure}[h!]
\begin{tabular}{cc}
\epsfig{file=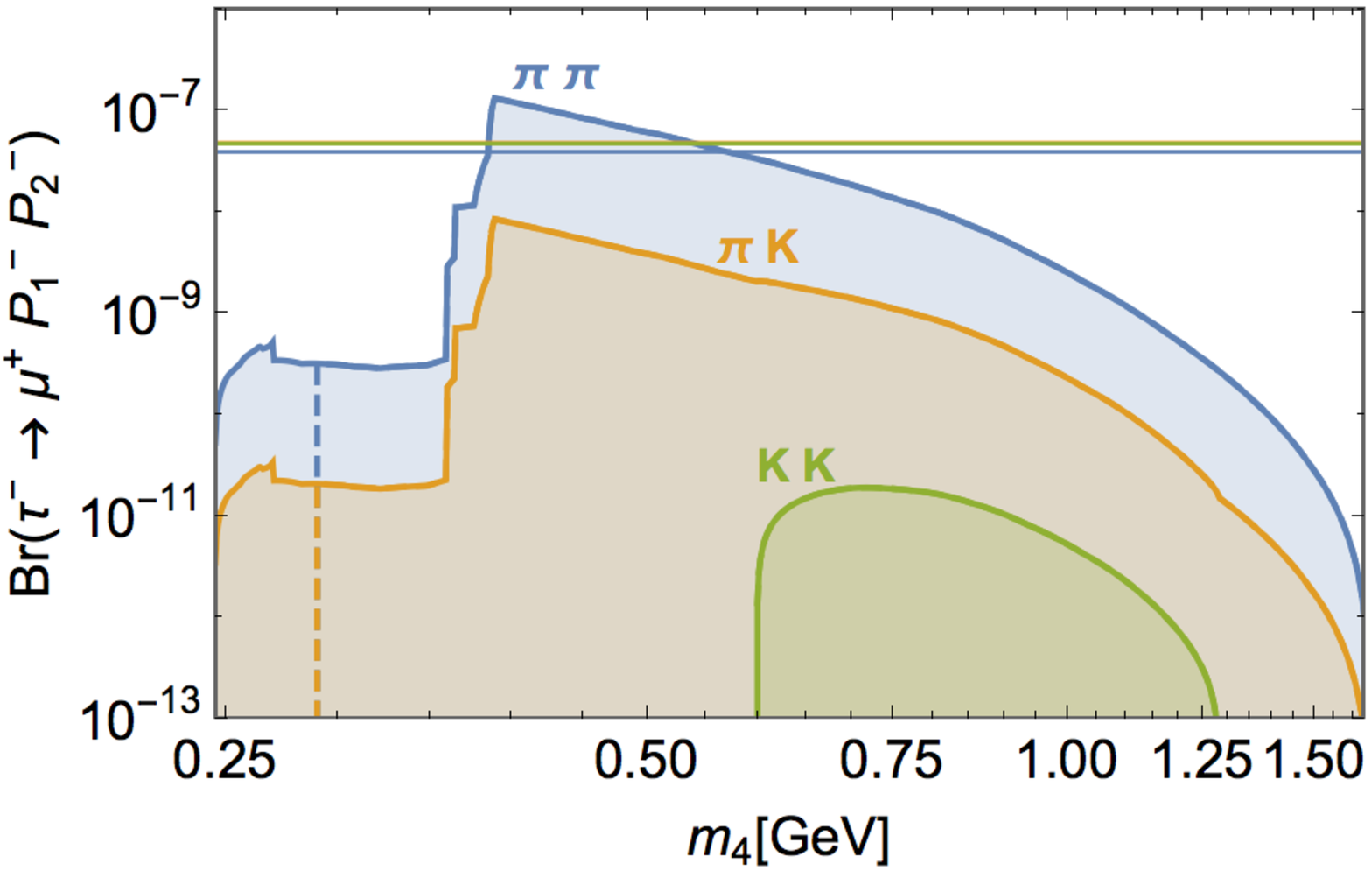, width=73mm} 
&
\raisebox{55mm}{\epsfig{file=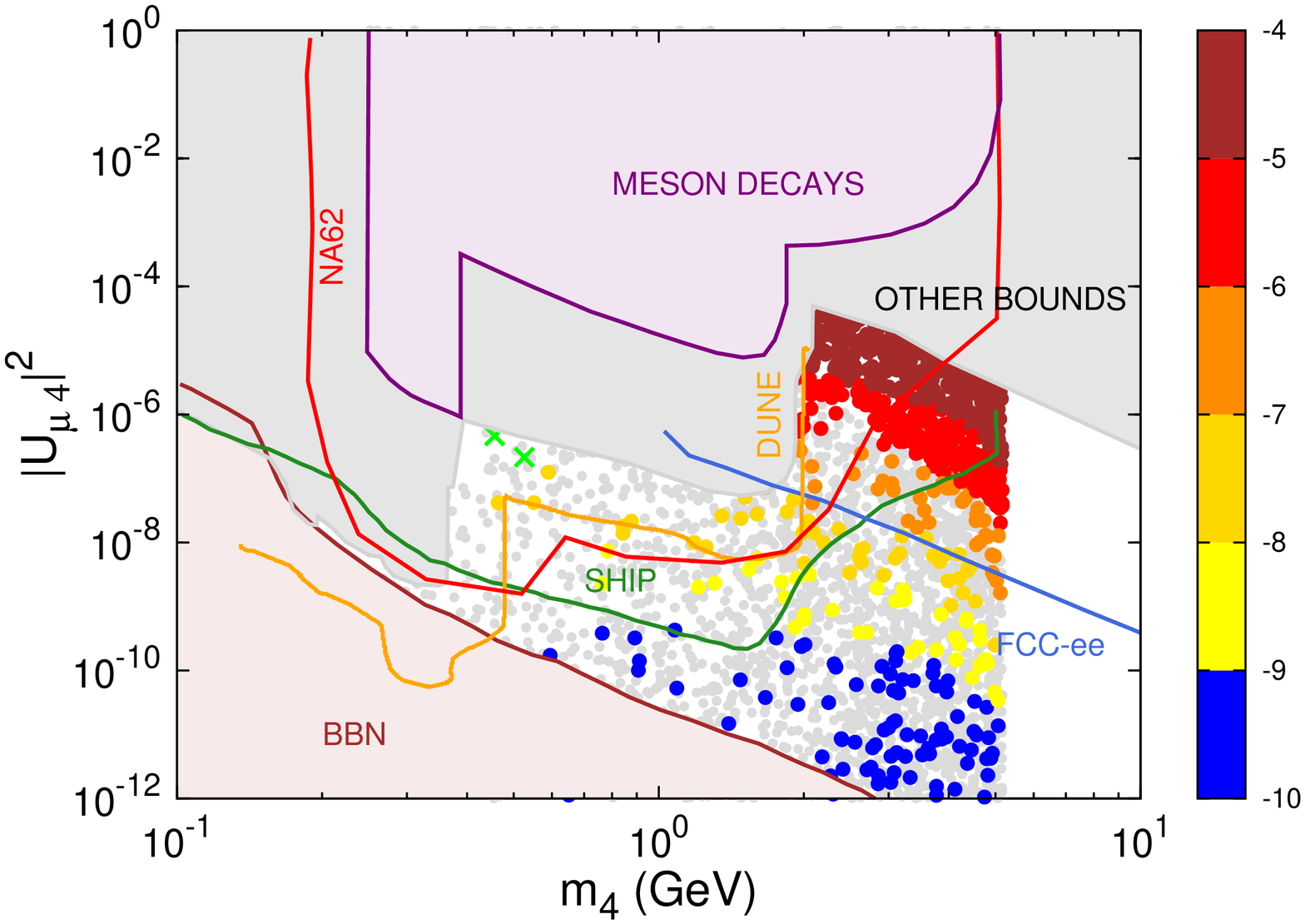, 
width=55mm, angle=-90}}
\end{tabular}
\caption{On the left, lines (surfaces) denoting the 
maximal (allowed) BR($\tau^- \to
\mu^+ P_1^- P_2^-$) vs. the heavy neutrino mass, $m_4$. 
On the right, predictions for the effective mass, $\log m_\nu^{\alpha
  \beta}$, in the $(|U_{\mu 4}|^2,m_4)$ plane, as derived from LNV
B-meson decays. Coloured surfaces and grey points denote excluded regimes.
From~\cite{Abada:2017jjx}.}\label{fig:LNV}
\end{figure}

Other than strongly impacting the prediction for $0\nu2\beta$ decays
(due to the augmented ranges for the effective mass in both normal and
inverted ordering schemes, a future observation can no longer be
straightforwardly associated with an inverted 
ordering~\cite{Giunti:2015wnd}, as visible on the right panel of
Fig.~\ref{fig:EDM.0nu2beta}), 
the sterile states can also be at the origin of LNV
semileptonic tau and meson decays. If produced on-shell, sterile neutrinos can
lead to a resonant enhancement of the LNV decay amplitudes, 
some processes already within experimental reach, as is the case of $\tau^- \to
\mu^+ \pi^- \pi^-$, or $K^+ \to \ell^+_\alpha \ell^+_\beta \pi^-$ 
(see left panel of Fig.~\ref{fig:LNV}). 
A comprehensive study of such
decays allows to infer bounds on all entries of a generalised
definition of the effective Majorana mass matrix: with the exception
of the $m_\nu^{\tau \tau}$ entry (whose bounds $\lesssim 10^{-2}$~GeV
strongly improve existing ones), all other entries are constrained to
lie below $\lesssim 10^{-4}$~GeV~\cite{Abada:2017jjx}. An example for
$m_\nu^{\mu \mu}$ is displayed on Fig.~\ref{fig:LNV}. 

\begin{figure}[h!]
\begin{tabular}{cc}
\hspace*{-5mm}
\epsfig{file=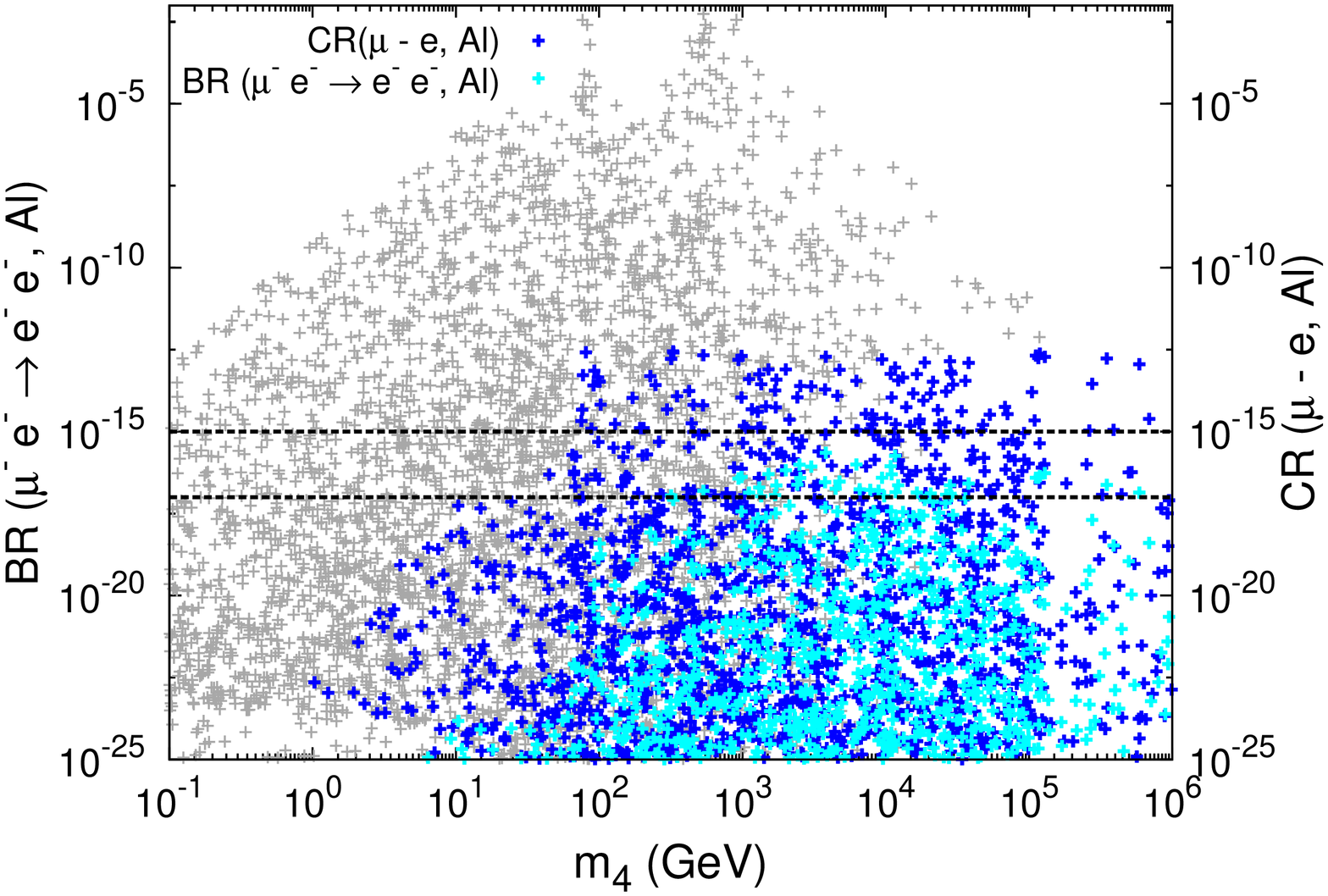, 
width=52mm, angle=-90} 
\hspace*{-2mm}&\hspace*{-2mm}
\epsfig{file=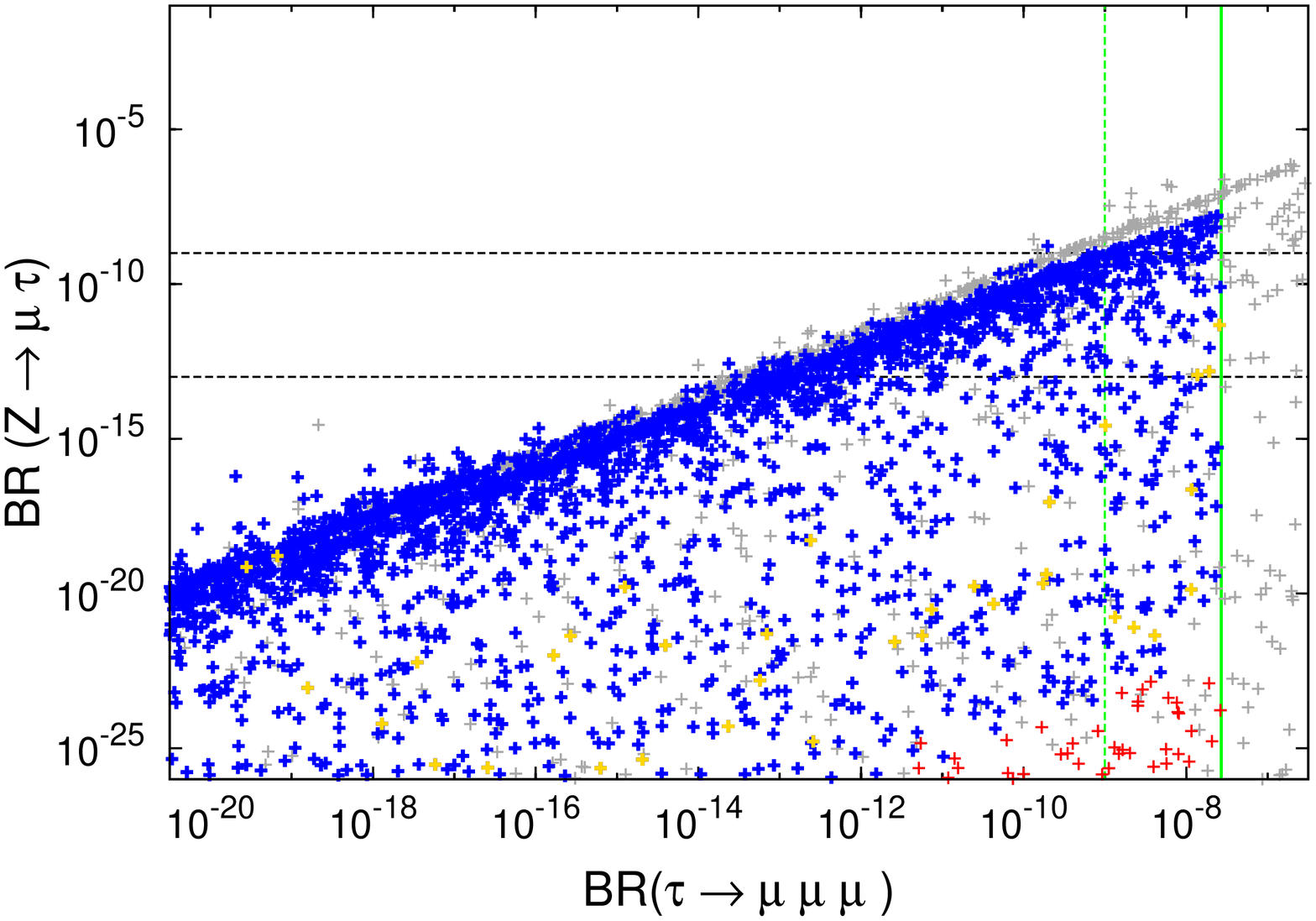, 
width=52mm, angle=-90}
\end{tabular}
\caption{BR($\mu^- e^- \to e^-e^-$, Al) (cyan, left axis) and 
CR($\mu-e $, Al) (dark blue, right axis) as a function of $m_4$. 
Grey points correspond to the violation of at least one experimental
bound; 
dashed horizontal lines denote the future sensitivity of COMET;
from~\cite{Abada:2015oba}. On the right, BR($Z \to \tau \mu$) vs. 
BR($\tau \to 3\mu$); blue (grey) points denote allowed (excluded)
regimes, and yellow points are associated with $0\nu 2\beta$ decays
within future
sensitivity; the upper (lower) horizontal line corresponds to the expected
sensitivity for a Linear Collider (FCC-ee), while vertical lines denote
current and future $\tau \to 3\mu$ sensitivities; 
from~\cite{Abada:2014cca}.}\label{fig:cLFV}
\end{figure}

Such a minimal construction also leads to important
contributions to cLFV observables: in the $e-\mu$ sector, neutrinoless
conversion in Nuclei (e.g. Aluminium) is one of the most sensitive observables
(although for heavier nuclei, the Coulomb-enhanced decay of a muonic
atom might be also competitive~\cite{Abada:2015oba}, see
Fig.~\ref{fig:cLFV}).  
For sterile states
heavier than the electroweak scale, three-body decays receive the
dominant contributions from $Z$-penguins, leading to a strong
correlation between the corresponding cLFV decays. In turn, this not
only allows to probe $\mu-\tau$ flavour violation 
beyond the reach of Belle II, but also
to explore this minimal SM extension at several 
frontiers~\cite{Abada:2014cca}, as displayed in Fig.~\ref{fig:cLFV}. 

\vspace*{-3mm}

\subsection{Mechanisms of neutrino mass generation}
\begin{figure}[h!]
\begin{tabular}{cc}
\hspace*{-5mm}
\epsfig{file=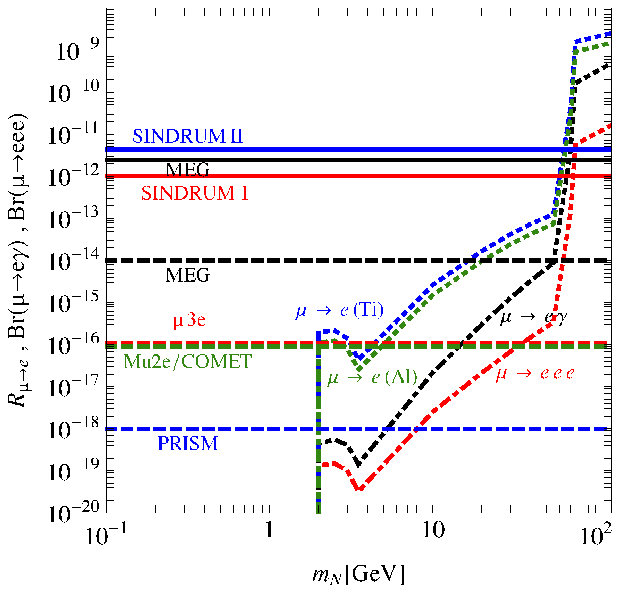, 
width=60mm} 
&
\raisebox{59mm}{\epsfig{file=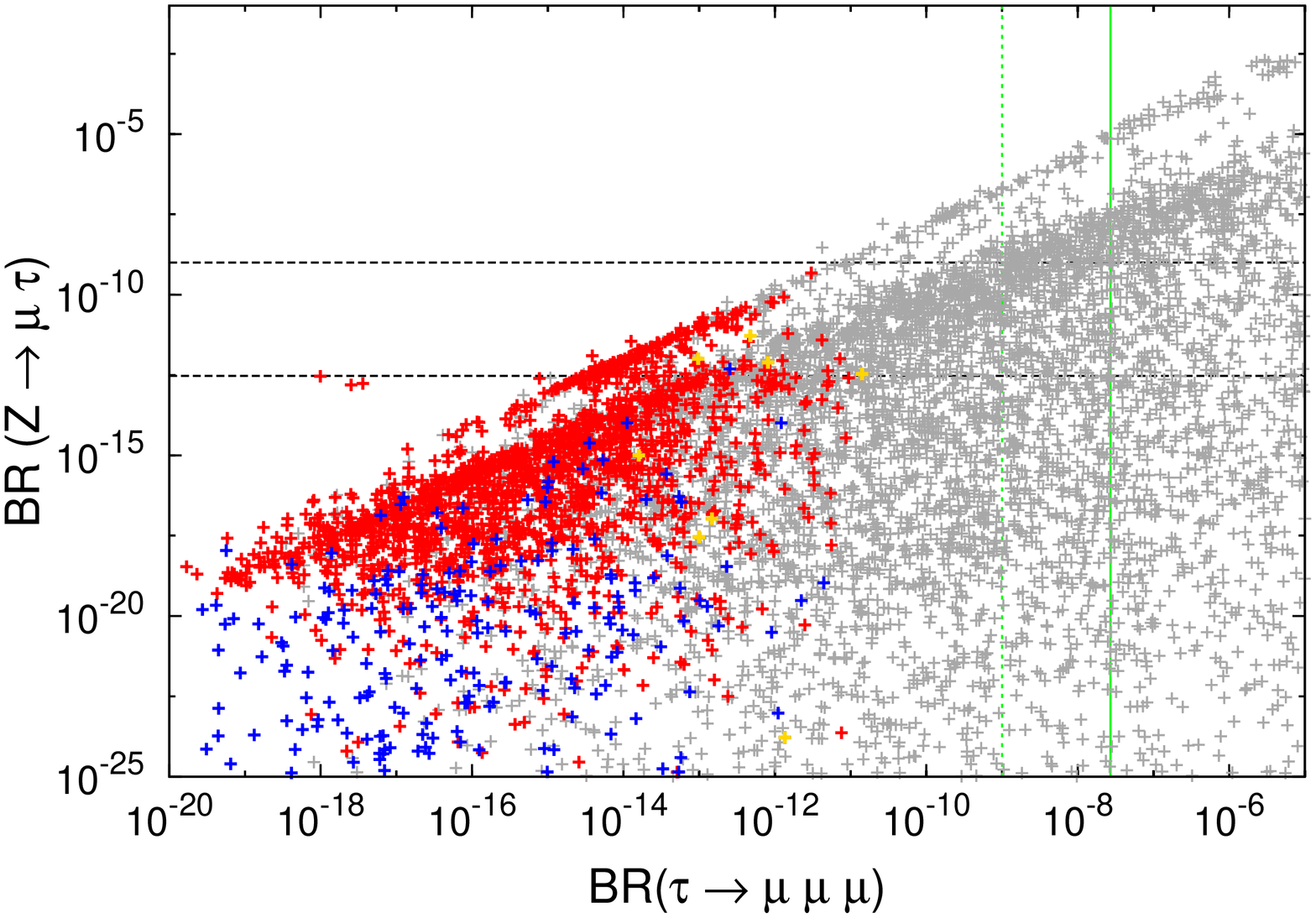, 
width=62mm, angle=-90}}
\end{tabular}
\caption{On the left, 
maximal allowed cLFV rates compatible with current searches
  in a low-scale seesaw; 
horizontal full (dashed) lines denote present (future) experimental 
sensitivity; from~\cite{Alonso:2012ji}.
On the right, BR($Z \to \tau \mu$) vs. 
BR($\tau \to 3\mu$) in a (3,3) ISS realisation; 
coloured (grey) points denote allowed (excluded)
regimes; the upper (lower) horizontal line corresponds to the expected
sensitivity for a Linear Collider (FCC-ee), while vertical lines denote
current and future $\tau \to 3\mu$
sensitivities; from~\cite{Abada:2014cca}. 
}\label{fig:models:cLFV}
\end{figure}

In its different realisations, the seesaw mechanism is perhaps one the
most appealing mechanisms of neutrino mass generation. Whether or not
a given seesaw realisation can be at the origin of a high-intensity
observable depends on the size of the Yukawa-like couplings, and
most importantly, on the scale of the new (heavy) mediators. 
The low-scale seesaw (and its variants) is an example of a type I
seesaw, whose mediators have non-negligible mixings with the active
neutrinos, and do not decouple. Not only can they give rise to
contributions to numerous cLFV observables (within future
sensitivity reach), but the high-intensity searches for the latter
allow to explore and constrain regions of the parameter space which
would be otherwise inaccessible~\cite{Alonso:2012ji} (see left panel
of Fig.~\ref{fig:models:cLFV}). 

Another phenomenologically and theoretically
appealing low-scale model of neutrino mass generation is the Inverse
Seesaw (ISS). In its (3,3) realisation, three sets of right-handed neutrinos
and extra sterile fermions are added to the SM content\footnote{Since
  in the ISS the light neutrino masses receive an extra suppression
  factor from the parameter which is at the origin of all LNV in the
  model ($\mu_X$), one can accommodate oscillation data with sizeable Yukawa
  couplings and a comparatively light NP scale by taking small values
  of $\mu_x$. The model remains theoretically natural, as in the limit
  $\mu_x \to 0$ one recovers lepton number conservation. 
}; the new states
do not decouple, leading to modified leptonic currents and extensive
contributions to many observables. For example, this is  
the case of cLFV muonic channels. However, and although $Z\to \tau
\mu$ decays are still within FCC-ee reach, $\tau \to 3 \mu$ lies
clearly beyond the reach of Belle II (cf. right panel
of Fig.~\ref{fig:models:cLFV}).
Although the ISS encompasses
several Dirac and Majorana CPV phases, having the heavy states forming
pseudo-Dirac pairs precludes significant contributions to 
lepton EDMs~\cite{Abada:2016awd}.
 
Due to the triplet nature of the mediators, both type II and type III
seesaws lead to very distinctive cLFV signatures. While in all type
I-like realisations, cLFV are higher order (loop) processes, in the
type II seesaw 3-body decays occur at tree-level; in the type III, both
3-body decays and coherent conversion in Nuclei are tree-level processes
(only radiative decays occur at loop level). 
By constructing ratios of observables, one can aim at disentangling
the different realisations: for example, 
BR($\mu\to e \gamma$)$/$BR($\mu \to 3 e$) $\sim 10^{-3} (\gtrsim 1)$
for type III (I); likewise
CR($\mu-e$, Ti)$/$BR($\mu \to e \gamma$) $\sim 10^{3} (\in [0.05 -5])$
for type III (II)~\cite{Hambye:2013jsa}.

\section{Embedding the seesaw in complete NP frameworks}
Aiming at addressing other observational (and theoretical) problems of
the SM, the seesaw can be embedded in larger, complete NP
frameworks, as is the case of supersymmetry (SUSY) or grand unified
theories (GUTs). 

\subsection{SUSY seesaw}
\begin{figure}[h!]
\begin{tabular}{cc}
\epsfig{file=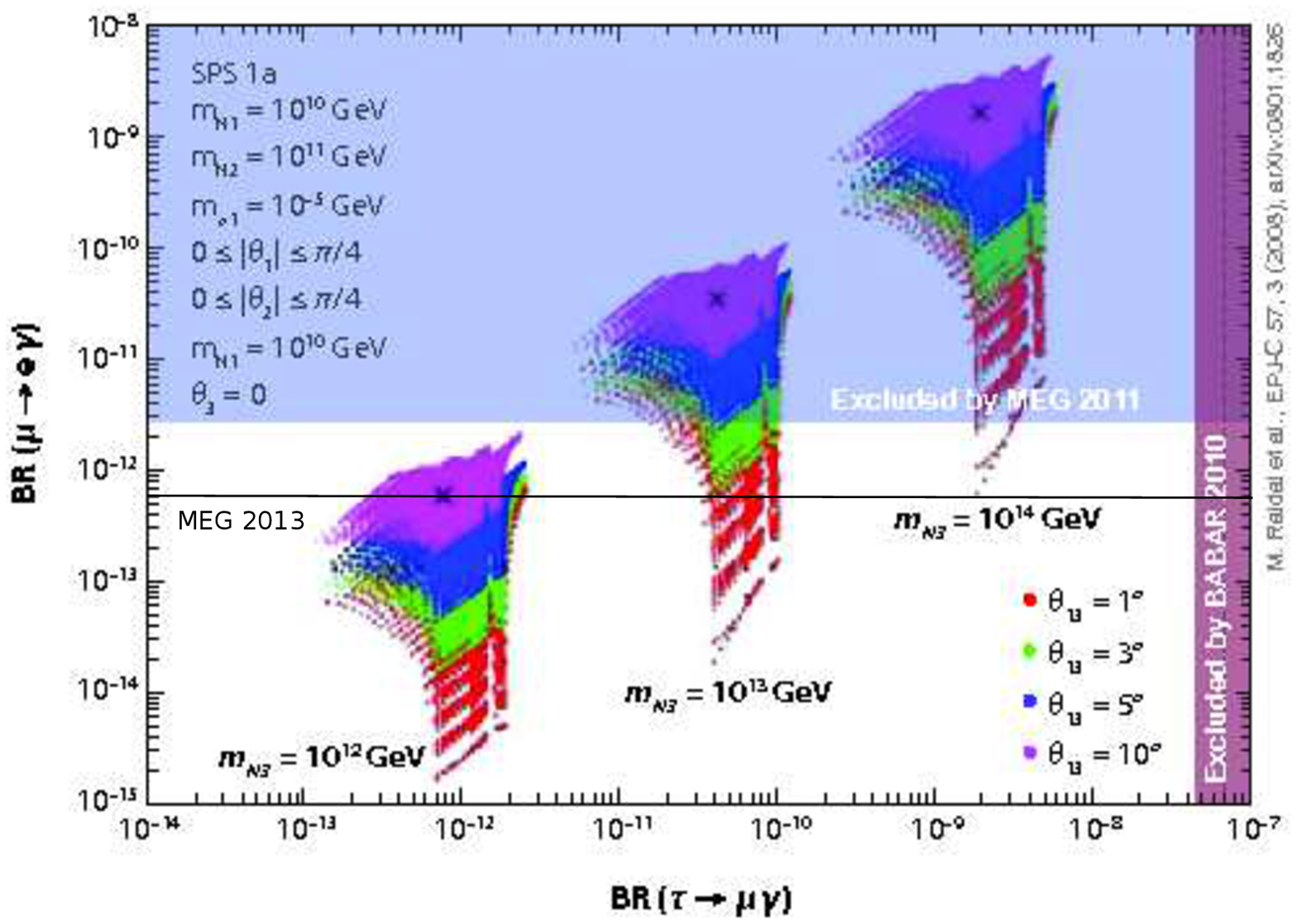, 
width=75mm} 
&
\epsfig{file=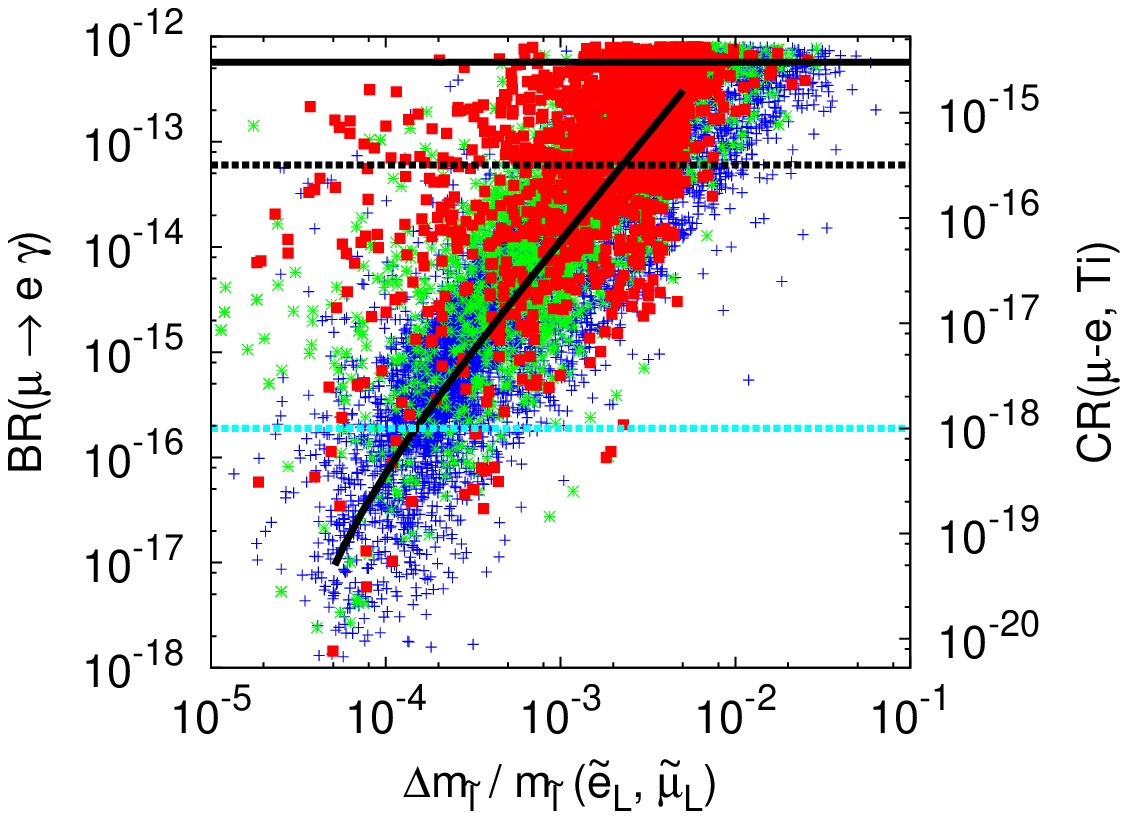, 
width=78mm}
\end{tabular}
\caption{Type I SUSY seesaw: on the left, correlation between 
BR($\mu \to e \gamma$) and BR($\tau \to \mu \gamma$) for different
seesaw scales, $M_R$; from~\cite{Antusch:2006vw}. 
On the right, BR($\mu \to e \gamma$) and CR($\mu-e$, Ti) vs. 
$\Delta_{\tilde \ell}/m_{\tilde \ell}(\tilde e_L, \tilde \mu_L)$, also for
different seesaw scales, $M_R$; 
from~\cite{Figueiredo:2013tea}.}\label{fig:models:cLFV.SUSY}
\end{figure}

The SUSY seesaw consists in the embedding of a (for example 
type I) seesaw in the
framework of otherwise flavour conserving SUSY models. Having a unique
source of LFV (the neutrino Yukawa couplings) implies that all
observables exhibit a strong degree of correlation. This is manifest
at low-energies in the strong synergy between $\mu \to e \gamma$ and
$\tau \to \mu \gamma$ decays, which remain tightly correlated
regardless of the typical SUSY spectrum or of the seesaw
scale~\cite{Antusch:2006vw} - see left panel of
Fig.~\ref{fig:models:cLFV.SUSY}.  
One can
further explore the synergy between low- and high-energy cLFV
observables (for instance new edges in dilepton mass
distributions, or relative mass differences between 
left-handed selectrons and smuons~\cite{Abada:2010kj})
to probe the SUSY seesaw hypothesis. Isolated cLFV manifestations
would disfavour the latter, while compatible ones  - as for example 
$\Delta m_{\tilde \ell}/m_{\tilde \ell}(\tilde e_L, \tilde \mu_L)
\gtrsim 0.5\%$ and an
observation of $\mu \to e \gamma$ at MEG -  would not only strengthen
it, but further hint on the seesaw scale ($M_R \sim
\mathcal{O}(10^{14}~\text{GeV})$)~\cite{Figueiredo:2013tea}, 
as visible on the right panel of Fig.~\ref{fig:models:cLFV.SUSY}.  

Increasing the degree of symmetry (be it in the form of extended gauge
symmetries, flavour ones, or gauge unification) reduces the
arbitrariness of the couplings, rendering the model more predictive,
and hence easier to test (and falsify). 
GUTs are particularly appealing and well-motivated theoretical
constructions: in addition to offering a common scheme for Yukawa
couplings, they can even relate observables in the lepton and quark sectors. 
In the simple case of a SU(5) type I SUSY seesaw, there is a strong
correlation between flavour violating observables - as well as CP
violating observables - in leptons and hadrons 
(see, for example~\cite{Calibbi:2009wk,Buras:2010pm}).
A second example of GUT-induced correlation of high-intensity
observables can be found in a leptogenesis motivated SO(10) type II
SUSY seesaw, which could be easily falsified by any future observation
of two low-energy cLFV processes.

\subsection{Further examples: vector-like leptons}
\begin{figure}[h!]
\begin{tabular}{cc}
\hspace*{5mm}
\epsfig{file=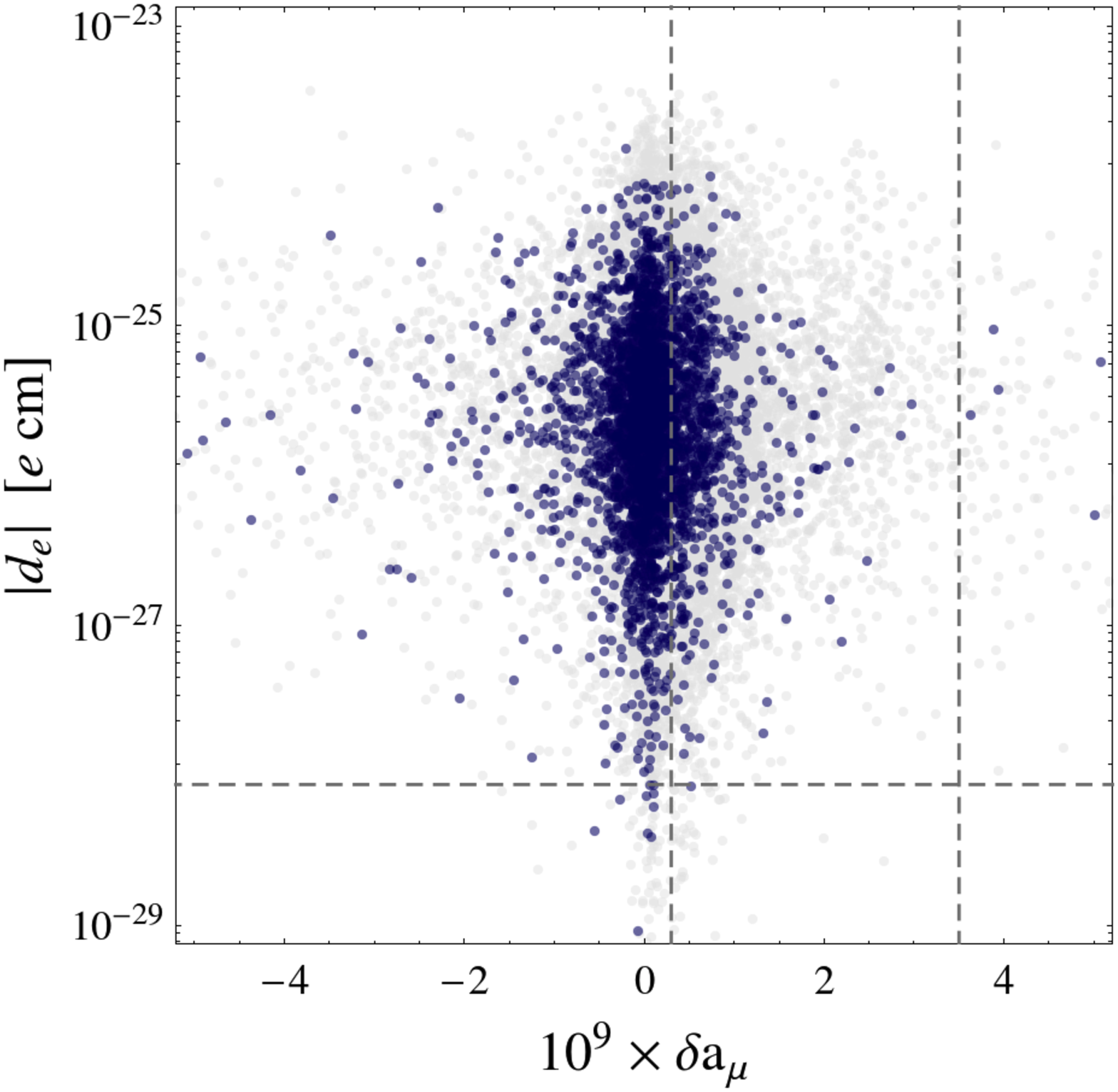, 
width=60mm} 
\hspace*{2mm}&\hspace*{2mm}
\epsfig{file=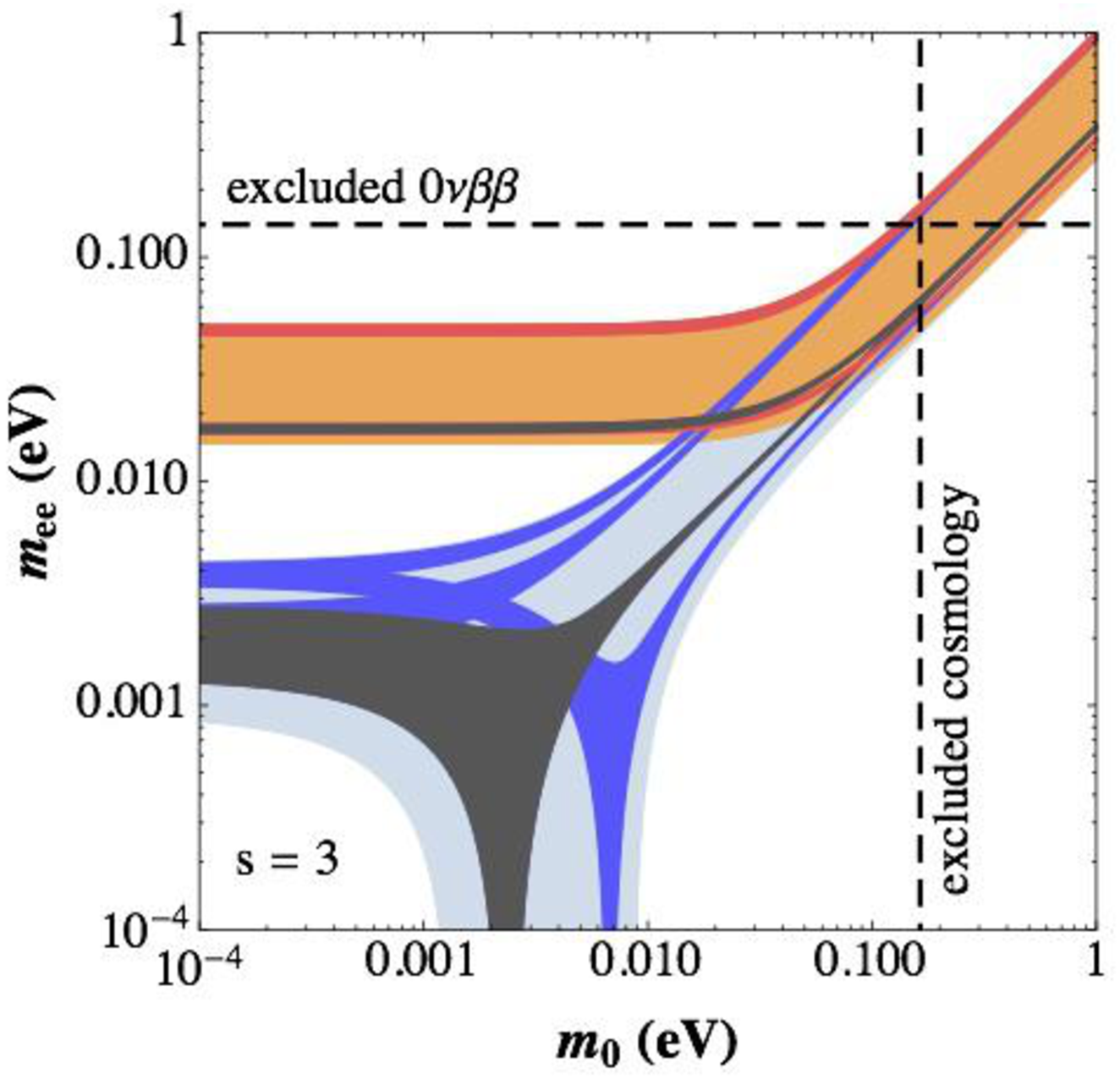, 
width=60mm}\hspace*{5mm}
\end{tabular}
\caption{On the left, correlation of $\delta a_\mu$
with the electron EDM for a model of vector-like leptons; 
grey points are ruled out by LHC, and 
dashed lines show the $2\sigma$ experimental region for $\delta a_\mu$, 
and the 90\% C.L. upper bound on $|d_e|$; from~\cite{Falkowski:2013jya}.
On the right, effective Majorana mass $m_{ee}$ as a function
of the lightest neutrino mass, for flavour groups $\Delta(3n^2)$ 
and specific classes of CP transformations; 
from~\cite{Hagedorn:2016lva}.
}\label{fig:further}
\end{figure}

Massive vector-like fermions are present in many well-motivated SM
extensions (as is the case of composite Higgs, warped extra
dimensions, ...). The prospects for cLFV (at high-intensities and in
Higgs decays) were addressed in~\cite{Falkowski:2013jya},
 for a generic set-up -
inspired by composite Higgs models - in which 3 generations of
vector-like left-handed ($L^V_i$) and right-handed ($E^V_i$) 
charged leptons were included. Neutrino masses can be obtained from
additional right-handed states, and the corresponding vector-like
partners. The contributions to cLFV observables (and lepton dipole
moments) turn out to be parametrised by a small set of couplings,
leading to correlated observables. For example one has 
BR($h\to \ell_i \ell_j$)$/$BR($\ell_i \to \ell_j \gamma$)
$\approx$ $4 \pi/3 \alpha$ 
BR($h\to \ell_i \ell_i$)$|_\text{SM}$$/$BR($\ell_i \to \ell_j \nu_i
\bar{\nu_j}$). 
Other than the latter synergy, 
a strong correlation between  EDMs and the muon 
anomalous magnetic moment ($\delta a_\mu$) was also
found. Interestingly, attempts to explain the current tension
between theory and observation in $(g-2)_\mu$ implies excessive
contributions to the electron EDM, almost leading to the exclusion of
the model - as can be seen from Fig.~\ref{fig:further} (left panel). 

\subsection{Further examples: flavour symmetries and massive neutrinos}
The flavour puzzle remains one of the most important open questions in
particle phyics. A possible way to address it, starting from first
principles, it to relate the flavour patterns (for example, the
texture of the Yukawa couplings) to the breaking of a flavour
symmetry $G_f$, 
continuous or discrete. This avenue has been extensively
explored in recent years, relying on very distinct approaches. The
only phenomenological caveat of certain constructions lies on the
difficulty of testing them - however, many realisations have
well-defined, peculiar signatures. We have discussed two illustrative 
examples:
(i) continuous flavour symmetry - minimal Abelian case, with 
$G_f =$U(1)$_{L_e + L_\mu} \times$U(1)$_{L_\tau}$ - leading 
to predictions of the BR($\mu \to e
\gamma$) correlated with the ordering scheme of the light neutrino
spectrum (for an example see~\cite{Deppisch:2012vj});
(ii) a discrete group based approach, with $G_f$ of the  
$\Delta (3 n^2)$ type, 
which predicts both lepton mixings as well as low- and high-scale
CPV phases~\cite{Hagedorn:2016lva} (see Fig.~\ref{fig:further}). 
Other than constraining predictions for
neutrinoless double beta decays, the latter
construction further leads to the
interplay of low-energy CP phases and a successful explanation of the
baryon asymmetry of the Universe from leptogenesis.  

\section{Overview}
While remaining one of the most important open questions in modern
particle physics, astrophysics and cosmology, 
neutrinos have proved to be true gateways to numerous new physics
phenomena. An extensive number of dedicated facilities is devoted to
searching for the latter, and the near future should see new data that will
hopefully clarify several points. 

Currently, a number of confirmed observations and several tensions
between experimental data and SM expectations suggests the need to
consider NP scenarios; interestingly, many of these tensions are
nested in lepton related-observables. We have briefly overviewed a
small subset of high-intensity observables\footnote{Further interesting
  observables include the Muonium system~\cite{Abada:2015oba} or
  ``in-flight cLFV conversion''~\cite{inflight}, among others.}, 
and the potential contributions of a few NP models, in particular of
realisations aiming at addressing the problem of neutrino mass
generation. We have also discussed several examples of how the synergy
between neutrino data and searches at the high-intensity frontier
might provide information on the underlying NP model of neutrino
mass generation.


\begin{thebibliography}{99}
\bibitem{NUFACT2017}
Proceedings of ``The 19th International Workshop on Neutrinos from
Accelerators (NUFACT2017)'', held 25-30 September 2017 at 
Uppsala University.

\bibitem{Patrignani:2016xqp}
C.~Patrignani {\it et al.} [Particle Data Group],
  Chin.\ Phys.\ C {\bf 40} (2016) no.10,  100001.

\bibitem{EFT:nufact17}
G.~M.~Pruna, these proceedings (WG4); A.~De Gouvea, 
these proceedings.

\bibitem{Abada:2015trh}
  A.~Abada and T.~Toma,
  JHEP {\bf 1602} (2016) 174.  

\bibitem{Giunti:2015wnd}
  C.~Giunti,
  Nucl.\ Phys.\ B {\bf 908} (2016) 336.

\bibitem{Abada:2017jjx}
  A.~Abada, V.~De Romeri, M.~Lucente, A.~M.~Teixeira and T.~Toma,
  ``Effective Majorana mass matrix from tau and pseudoscalar meson
  lepton number violating decays,'' 
  arXiv:1712.03984 [hep-ph].

\bibitem{Abada:2015oba}
  A.~Abada, V.~De Romeri and A.~M.~Teixeira,
  JHEP {\bf 1602} (2016) 083. 

\bibitem{Abada:2014cca}
  A.~Abada, V.~De Romeri, S.~Monteil, J.~Orloff and A.~M.~Teixeira,
  JHEP {\bf 1504} (2015) 051.

\bibitem{Alonso:2012ji}
  R.~Alonso et al, 
  JHEP {\bf 1301} (2013) 118.

\bibitem{Abada:2016awd}
  A.~Abada and T.~Toma,
  JHEP {\bf 1608} (2016) 079.

\bibitem{Hambye:2013jsa}
  T.~Hambye,
  Nucl.\ Phys.\ Proc.\ Suppl.\  {\bf 248-250} (2014) 13.

\bibitem{Antusch:2006vw}
  S.~Antusch, E.~Arganda, M.~J.~Herrero and A.~M.~Teixeira,
  JHEP {\bf 0611} (2006) 090.

\bibitem{Abada:2010kj}
  A.~Abada, A.~J.~R.~Figueiredo, J.~C.~Romao and A.~M.~Teixeira,
  JHEP {\bf 1010} (2010) 104.

\bibitem{Figueiredo:2013tea}
  A.~J.~R.~Figueiredo and A.~M.~Teixeira,
  JHEP {\bf 1401} (2014) 015.

\bibitem{Calibbi:2009wk}
  L.~Calibbi et al, 
  JHEP {\bf 0912} (2009) 057.

\bibitem{Buras:2010pm}
  A.~J.~Buras, M.~Nagai and P.~Paradisi,
  JHEP {\bf 1105} (2011) 005.

\bibitem{Falkowski:2013jya}
  A.~Falkowski, D.~M.~Straub and A.~Vicente,
  JHEP {\bf 1405} (2014) 092.

\bibitem{Deppisch:2012vj}
  F.~F.~Deppisch,
  Fortsch.\ Phys.\  {\bf 61} (2013) 622.

\bibitem{Hagedorn:2016lva}
  C.~Hagedorn and E.~Molinaro,
  Nucl.\ Phys.\ B {\bf 919} (2017) 404.

\bibitem{inflight}
M. Yamanaka, these proceedings; see also 
  A.~Abada, V.~De Romeri, J.~Orloff and A.~M.~Teixeira,
  Eur.\ Phys.\ J.\ C {\bf 77} (2017) no.5,  304.


\end{thebibliography}
\end{document}